\documentclass[twocolumn,showpacs,aps,prd]{revtex4-1}
\usepackage{graphicx} 

\usepackage{amsmath}
\usepackage{subfigure}
\usepackage{epstopdf}
\usepackage{multirow}
\usepackage{xspace}
\usepackage{rotating}
\usepackage{longtable}
\usepackage{multirow}
\usepackage[breaklinks=true]{hyperref}
\hypersetup{
  colorlinks=true,
  linkcolor=blue,
  citecolor=blue,
  urlcolor=blue
}

\graphicspath{{./fig/}}
\usepackage{array,tabularx,epsfig,mathrsfs,graphicx,rotating}
\usepackage{ifthen}
\usepackage{amsfonts}
\newcommand{\beq}{\begin{equation}}
\newcommand{\eeq}{\end{equation}}

\chardef\til=126

\newcommand{\zprime}{{\,\mathrm{Z^{0\prime}}}}
\newcommand{\gkk}{{\,\mathrm{g_{KK}}}}

\newcommand{\abb}{\mathrm{ab^{-1}}}

\begin{document}

\title{\boldmath
Sensitivity to new high-mass states decaying to  $t\bar{t}$ at a 100~TeV collider
}
\author{B.~Auerbach, S.~Chekanov,  J.~Love, J.~Proudfoot}
\affiliation{
HEP Division, Argonne National Laboratory,
9700 S.Cass Avenue,  
Argonne, IL 60439
USA
}

\author{A.~V.~Kotwal}

\affiliation{
Fermi National Accelerator Laboratory and Department of Physics, Duke University, USA 
}

\date{\today}


\begin{abstract}
We discuss the sensitivity of a 100 TeV $pp$ collider to heavy particles decaying to top-antitop $(t\bar{t})$ final states.
This center-of-mass energy, together with  an integrated luminosity of $10$~ab$^{-1}$,  can produce heavy particles
in the mass range of several tens of teraelectronvolts (TeV). 
A Monte Carlo study has been performed using boosted-top techniques to reduce QCD background 
for the reconstruction of heavy particles with masses in the  range of 8--20~TeV, 
and various widths.     
In particular, we have studied two models that predict heavy states, a model with   
an extra gauge boson ($\zprime$) and with a Kaluza-Klein (KK) excitation of the gluon ($\gkk$). 
We estimate the sensitive values of $\sigma \times$Br of about 2 (4)~fb for $\zprime$ ($\gkk$), with a
corresponding mass reach of 13 (20)~TeV. 
\end{abstract}

\pacs{12.38.Qk,  14.80.Rt}

\maketitle


\section{Introduction}

The top ($t$) quark, the highest-mass particle so far observed, 
 may  be  closely related to new physics beyond the TeV scale due to possible strong 
coupling to new, more massive particles.
Therefore, final states containing top quarks are motivated for  
exploring physics opportunities of a 100~TeV proton-proton collider, which is capable of 
producing  exotic particles with masses close to or above 10~TeV.  Due to 
the significant Lorentz boost from such massive resonance decays, 
the top-quark decay products often overlap partially or completely. 
Such decay products cannot be reconstructed as separate objects using ``resolved'' techniques which identify
individual objects as jets (from the hadronisation of quarks) or high-$p_T$ leptons.  

This study focuses on searches for new, heavy particles that 
decay to $t\bar{t}$, a generic consequence of many Beyond-the-Standard Model (SM) theories. 
For example,
the existence of such particles was discussed in the framework of a
generic Randall-Sundrum  model \cite{PhysRevLett.83.3370}.
This model predicts  a number of heavy particles, such
as an extra gauge boson (see the review \cite{Langacker:2008yv})
or Kaluza-Klein (KK) excitations of the gluon \cite{Lillie:2007yh}.
In the following, we denote such particles with the symbols $\zprime$ or $\gkk$, respectively.
We will discuss the  simulation of $\zprime/\gkk$ processes in the following sections.

Jet substructure and jet shapes are often discussed as  useful tools to
distinguish events produced by the standard QCD processes from those containing
jets arising from decays of multi-TeV  particles.
Such methods have been employed by the LHC experiments to increase sensitivity to high-mass states decaying to $t\bar{t}$~\cite{CMS:2011bqa,Aad:2012dpa}.
There is a diverse phenomenology of such approaches~\cite{Agashe:2006hk,Lillie:2007yh,Butterworth:2007ke,Almeida:2008tp,Almeida:2008yp,Kaplan:2008ie,Brooijmans:2008,Butterworth:2009qa,Ellis:2009su,Ellis:2009me,ATL-PHYS-PUB-2009-081,CMS-PAS-JME-09-001,Thaler:2010tr,Chekanov:2010vc,Chekanov:2010gv,Almeida:2010pa,Hackstein:2010wk}. 
A comparison of these techniques has been performed by both LHC experiments~\cite{TheATLAScollaboration:2013qia,CMS:2014fya} 
and is outside the scope of this paper. 

The goal of this paper is to understand the physics reach for heavy particles decaying to $t\bar{t}$ 
that may be produced at a  100~TeV $pp$ collider, assuming an integrated luminosity of $10$~$\abb$. 
As mentioned above, the identification of such particles is challenging due to the large boost of the top quarks, making signal events almost indistinguishable from SM two-jet events
 such as QCD dijet and $W/Z$  processes.

\begin{figure}[htp]
\centering
\includegraphics[scale=0.4]{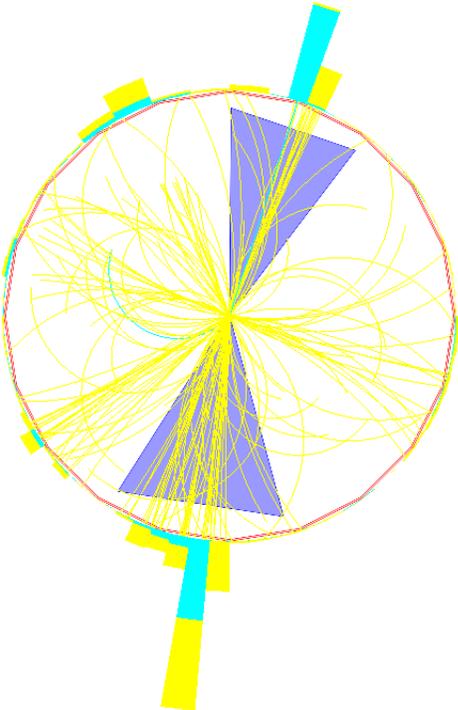}
\caption{An event display of a typical $\zprime$($M=10$~TeV) decay to $t\bar{t}$ at
a $100$ TeV $pp$ collider. 
Two jets with transverse momenta above 3~TeV are shown with the dark blue cones. The jets are 
reconstructed using the anti-$k_T$ algorithm \cite{Cacciari:2008gp}
with  a distance parameter of 0.5 using the {\sc FastJet} package~\cite{fastjet}.
Yellow lines show charged hadrons and light-blue lines show contributions from electrons.
}
\label{View1}
\end{figure}

The production cross section of the heavy particle is  model-dependent. For a given model, higher-order QCD corrections to the production cross section 
can be as large as 100\% \cite{Gao:2010bb}.
Therefore, our focus is to extract the sensitivity to a generic heavy particle decaying
to $t\bar{t}$ using boosted techniques, and to illustrate limitations that can be faced in the 20~TeV mass region.

This analysis uses all decay channels for top quarks. In the case of the 
leptonic decay of one top (or anti-top) quark, we will correct for
the missing neutrino as explained in Sect.~\ref{sec:dijet}. The contribution from fully-leptonic decays of both quarks can be ignored
since such events typically do not pass the selection requirements. 

An example of the decay $\zprime \to t\bar{t}$ is shown in Fig.~\ref{View1}.
It shows two jets with transverse momenta above 3~TeV 
that originate from boosted top (anti-top) quarks. The mass of the $\zprime$ boson was set to 10~TeV.  
The event display was created using the {\sc Delphes} 
fast simulation \cite{deFavereau:2013fsa} and the Snowmass detector setup \cite{Anderson:2013kxz}.
The event, generated with the {\sc Pythia8} Monte Carlo generator \cite{Sjostrand:2006za},
was taken from the {\sc HepSim} repository \cite{Chekanov:2014fga}.
The figure illustrates that both top jets can be reconstructed using the anti-$k_T$ algorithm \cite{Cacciari:2008gp}
with a distance parameter of 0.5.   A complete color version of this event is shown in the Appendix.

This study is performed without simulation of the detector response. The inclusion of a realistic detector
simulation is an important component for future studies. The only assumption is the $b$-tagging performance that will be discussed below. Since this is one of the most important
discriminating variables, we believe that the inclusion of a realistic MC simulation with a similar $b$-tagging performance (efficiency and mistag rates)
will not change significantly the results of this analysis.

\section{Monte Carlo simulations}

The analysis was performed using the {\sc Pythia8}~\cite{Sjostrand:2006za}, 
{\sc Herwig++}~\cite{Bahr:2008pv}  and {\sc Madgraph5}~\cite{Alwall:2011uj} 
MC models with the default parameter settings.
The MSTW2008lo68cl \cite{Martin:2009iq} parton density function (PDF) set was used.

The SM $t\bar{t}$ predictions were performed at next-to-leading order (NLO) in QCD with the {\sc Madgraph5} program.
 The transverse momenta ($p_T$) of the top quarks were required to be $ p_T > 2.5$~TeV to increase the efficiency for event generation.
 The $t\bar{t}$ cross sections were calculated at leading-order (LO). We have found that the ratio  
 of  the NLO to LO cross sections is a factor $\sim 3$ for $p_T(t)>2.5$~TeV. This relatively large 
``k-factor'' for high-$p_T$ top-quark events was independently checked using other calculations \cite{discussion}. 

Dijet QCD background was simulated with {\sc Pythia8}~\cite{Sjostrand:2006za}.
This model is used to generate all $2\rightarrow 2$ quark and gluon processes, including $b$-quark pair production, 
except for $t\bar{t}$ production. 
The MC inclusive-jet cross section was scaled to match the NLO prediction
estimated with the NLOjet++ program \cite{Catani:1996vz,Nagy:2003tz} and the MSTW2008nlo68cl PDF set \cite{Martin:2009iq}. 
The estimated $k$-factor used to multiply the {\sc Pythia8} (or {\sc Herwig++}) cross section is $1.23$.

The {\sc Herwig++} \cite{Bahr:2008pv} generator is used to cross-check the  {\sc Pythia8} background sample. 
{\sc Herwig++} includes a simulation of soft $W/Z$ boson radiation within the parton shower, 
thus this generator provides an alternate event sample for background processes.

\begin{figure}[htp]
\centering
\includegraphics[scale=0.4]{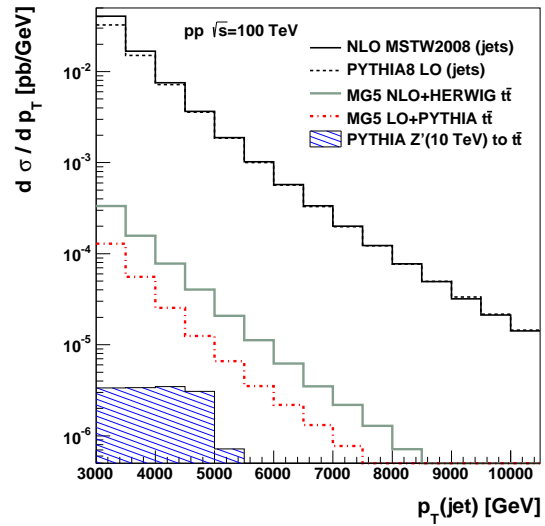}
\caption{Inclusive jet cross sections  as a function of the leading jet $p_T$
for several MC simulations used in this analysis.
}
\label{h_pt}
\end{figure}

 {\sc Pythia8} is also used to produce event samples with $W$- and $Z$-boson processes. They include double-boson production and
$W/Z$+jet production calculated using LO QCD. 
After signal selection, contributions from these background processes are negligible.
Therefore, we did not attempt to use simulations of 
multi-parton hard processes included in  {\sc Alpgen}~\cite{Mangano:2002ea} or {\sc Blackhat}~\cite{Bern:2012my},  that typically
lead to larger cross sections.

Events with heavy particles decaying to $t\bar{t}$ were generated using two models: a model of  
an extra gauge boson, $\zprime$, and a Randall-Sundrum Kaluza-Klein  gluon, $\gkk$. 
$\zprime$ boson events were generated using the pure $q \bar{q} \to \zprime$ production process (i.e. ignoring interference with SM processes) 
calculated using LO QCD as implemented in {\sc Pythia8}. 
The detailed description of such models and their default parameters can be found in the {\sc Pythia8} manual~\cite{Sjostrand:2006za}.
The KK gluons are simulated using the model in Ref.~\cite{Ask:2011zs}, with the process $q \bar{q} \to \gkk$ and also ignoring interference terms
with SM processes.
Although the models generate the boosted $t\bar{t}$ topology similarly, 
the decay widths and the production rates of $\zprime$ and $\gkk$ are different. 
The width of the $\zprime$ boson was set to $\Gamma/M=3\%$, while the width
of $\gkk$ is substantially larger, $\Gamma/M=16\%$. The $\gkk$ production rate
is more than a factor of ten larger than that of $\zprime$ boson. 

The production rates of $\zprime$ and $\gkk$ were calculated  using  LO QCD as
NLO corrections  are  not known. Recently, it was shown that
NLO contributions to $\zprime$  production  can be as large as 100\% for certain models and masses \cite{Gao:2010bb}. 
In the following, we  multiply the $\zprime$ cross section by 1.3 as was done at lower-energy
colliders, in order to maintain consistency with previous studies. 

All Monte Carlo samples used in this study are available
from the {\sc HepSim} public repository \cite{Chekanov:2014fga}
that stores theoretical predictions in the {\sc ProMC} file format
\cite{2013arXiv1311.1229C,2013arXiv1306.6675C}. The  samples
were analyzed  with a C++/ROOT program \cite{root}.
To illustrate the scale of the computation, the largest MC dataset analysed from {\sc Hepsim} 
was the dijet background sample with 0.4 billion $pp$ collision events.  
The jets were reconstructed with the anti-$k_T$ algorithm \cite{Cacciari:2008gp}
with  a distance parameter of 0.5 using the {\sc FastJet} package~\cite{fastjet}.
Jets are selected using the requirements $p_T (\mathrm{jet})>2.8$~TeV and $|\eta(\mathrm{jet})|<3$.
For jet clustering, stable particles are selected
if their mean lifetimes are larger than $3\cdot 10^{-11}$ seconds.
Neutrinos are excluded from consideration in jet clustering.

As discussed above, no simulation of the detector response was applied.  
This  analysis focuses  on the potential of a future
proton collider at 100~TeV, taking into account limitations arising from statistics, approximate SM background rates, 
and the effect of background-rejection methods.
The performance of $b$-tagging is assumed to be similar to the LHC experiments at lower $b$-quark momenta. 
A detailed detector geometry requires significant studies that are beyond the scope 
of this analysis. The study presented here can be useful for informing the design of detectors  and for model builders.  

Figure~\ref{h_pt} shows the inclusive jet cross sections at NLO and LO 
for SM (``background'') processes.  Since the predictions of {\sc Pythia8} and {\sc Herwig++} are similar, 
only {\sc Pythia8} is shown, which is also used to estimate the 
QCD dijet background. The {\sc Pythia8} cross section was scaled by the $k$-factor of 1.23 using NLOjet++  
as explained above.  We use the $t\bar{t}$ cross section calculated at NLO QCD as discussed above.
Note that there is a large difference between the LO and NLO calculations shown in Figure~\ref{h_pt}, which explains the  large $k$-factor discussed earlier. 

\begin{figure}[htp]
\centering
\includegraphics[scale=0.4]{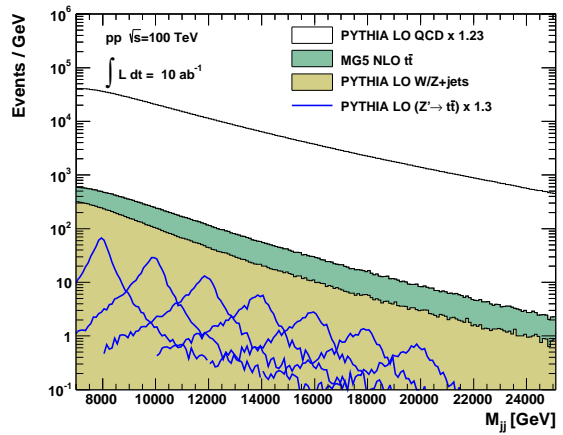}
\caption{Distributions of dijet mass at a 100 TeV collider for 10~$\abb$ for background processes,  together with expected $\zprime \to t\bar{t}$ signals
generated at different masses. All SM predictions are shown as the stacked histogram.
The white histogram is 
dominated by light-flavor dijet events from {\sc Pythia}.
The signal predictions (shown with the blue lines) for  $\zprime \to t\bar{t}$  are generated at LO QCD and scaled by the NLO-QCD $k$-factor of 1.3  as
explained in the text. 
}
\label{h_jetjet_zprime0}
\end{figure}

\begin{figure}[htp]
\centering
\includegraphics[scale=0.4]{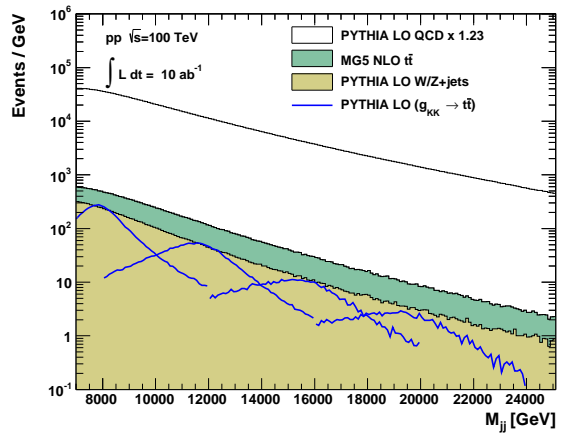}
\caption{Same as Fig.~\ref{h_jetjet_zprime0}, but with the expectation for the $\gkk \to t\bar{t}$ signal
generated at different masses.
All SM predictions are shown as the stacked histogram.
The signal predictions for  $\gkk$  are performed using LO QCD (the blue lines).
}
\label{h_jetjet_kkgluon0}
\end{figure}

\section{Dijet mass distributions}
\label{sec:dijet}

This analysis uses dijet mass distributions to extract the sensitivity to $t\bar{t}$ resonances above 8~TeV.
The decay products of top quarks from such resonances are well-contained within the jets, 
assuming the jet distance parameter $R=0.5$~\footnote{Recent studies 
based on similar top-quark transverse momenta and anti-$k_{T}$ jets with $R=0.5$ were shown in Ref.~\cite{Calkins:2013ega}}. 

The dijet invariant-mass distribution is an obvious choice for searches  for $t\bar{t}$ resonances using the traditional ``bump hunt'' procedure.
The major backgrounds arise due to QCD dijet production, SM $t\bar{t}$ production, and $W/Z$ production. The QCD dijet production 
 includes all $2 \to 2$  light-flavor quark and gluon processes and prompt $b$-quark pair production. 

In the case of fully hadronic decays, the reconstruction of two top-jets from $\zprime/\gkk$ decays will lead to a bump in the dijet invariant-mass distribution near the nominal mass. 
When both top (anti-top) quarks decay hadronically, we expect  two back-to-back jets that contain decay products of each quark.
When one top quark decays leptonically, there is an imbalance in the jet $p_T$'s due to the unmeasured neutrino.
To take into account the missing $p_T$ for the reconstruction of the
$t\bar{t}$ invariant mass, we add the missing $p_T$ to the $p_T$ of sub-leading jet. 
This {\em ad hoc} method improves the di-top mass resolution for semi-leptonic $t\bar{t}$  decays.

Figures~\ref{h_jetjet_zprime0} and~\ref{h_jetjet_kkgluon0} show the dijet invariant mass distributions for SM and BSM heavy particle processes, $\zprime$ and $\gkk$, generated with different 
 masses. 
The blue line shows the $\zprime/\gkk \to t\bar{t}$ contributions where the heavy particles
were generated with masses ranging from 8 to 20~TeV.
As expected, the $\zprime$ and $\gkk$ signals have a Gaussian-like shape corresponding to fully-merged
decays of top quarks. 
The dijet-mass distribution is dominated by the SM light-flavor dijet processes, therefore, an observation of a bump
with the  $\zprime/\gkk$ cross section is extremely challenging.

\begin{figure}[htp]
\centering
\includegraphics[scale=0.4]{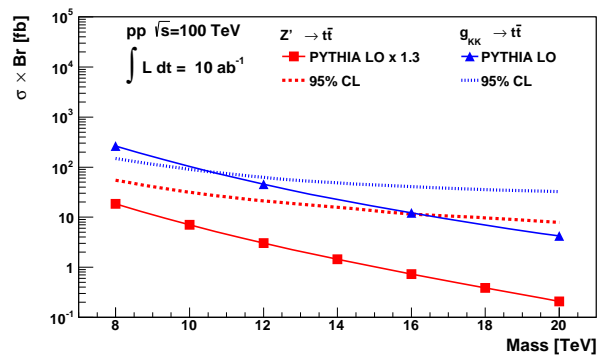}
\caption{The theoretical $\sigma \times$Br for $\zprime \to t\bar{t}$ and $\gkk \to t\bar{t}$ processes,  
 together with the $2\sigma$ sensitivity values  for 10~$\abb$ of integrated luminosity. Cuts designed to increase the 
S/B ratio have not been made. These 95\% CL sensitivity estimates ignore detector reconstruction effects and systematic uncertainties. 
}
\label{exclu_plot0}
\end{figure}

Figure~\ref{exclu_plot0} shows the theoretical cross sections ($\sigma$) times the branching ratio (Br) 
for the $\zprime \to t\bar{t}$ ($\gkk \to t\bar{t}$) process as a function of resonance mass. Also shown is  the value of the $\sigma \times$Br 
needed, at each resonance mass,  for excluding the background-only hypothesis at the $95\%$ confidence level (CL). This ``2$\sigma$ evidence'' value of $\sigma \times$Br for the signal is calculated  
 using the $CL_b$ method as implemented 
in the {\sc MClimit} program \cite{Junk:1999kv} and includes statistical uncertainties on  background and signal. 
For a counting experiment, $CL_b$ specifies the probability that the generated number of events from the background-only hypothesis   
 is smaller than or equal to the observed number, $CL_b=P_b(N\leq N_{\rm obs})$. 
In the presence of signal, poor compatibility of the observation with the background-only hypothesis is indicated by $CL_{b}$ being close to one.
The ``2$\sigma$'' sensitivity values shown in Fig.~\ref{exclu_plot0} represent the signal $\sigma \times$~Br for which the background-only hypothesis is excluded at the $95\%$ CL (i.e. $CL_b > 0.95$).   
It can be seen that the experimental signal sensitivity is substantially 
worse than the $\sigma \times$Br for the $\zprime$ model. 
Figure~\ref{exclu_plot0} shows that the $\gkk$ signal can be observed at 2$\sigma$ for masses below $10$~TeV, given the large signal statistics
in this region. However, the signal-over-background (S/B) ratio in this mass region is only $0.2\%$, thus systematic uncertainties should be understood
below this level in order to observe the $\gkk$ signal.  
For realistic measurements, such a level of systematic precision is difficult to achieve.

\begin{figure}[htp]
\centering
\includegraphics[scale=0.4]{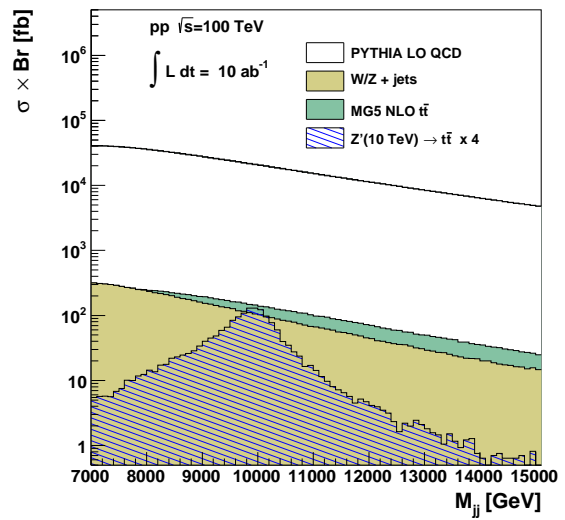}
\caption{An illustration of a  $2 \sigma$ ``evidence'' scenario assuming a  $\zprime$ boson with the mass of 10 TeV.
The $\zprime$  cross section  
 was scaled by the factor of 4 in order to  
exclude the background-only hypothesis with $95\%$ confidence. 
Similar to Fig.~\ref{h_jetjet_zprime0}, all background histograms are stacked. The scaled signal histogram is shown
 separately and also added to the stack. 
}
\label{h_jetjet_all}
\end{figure}

Figure~\ref{h_jetjet_all} illustrates the scenario needed to exclude the background-only hypothesis with $95\%$ confidence, by scaling the $\zprime$ signal by a factor that is needed for such exclusion. 
As before, no  requirements to reduce the QCD background are made. 
This figure shows that even if the  $\zprime$  cross section is
scaled by a factor of 4, the signal-over-background ratio near 10 TeV is only $0.3\%$, thus an  observation of such a state
will require understanding of systematic uncertainties below this level.
Experimental realization of such systematic precision is very challenging.

\section{Discriminating variables}

In order to reduce QCD background and increase sensitivity to a $t\bar{t}$ resonance,
we apply several requirements based on jet mass, sub-jet information and $b$-tagging. 

\begin{figure}[htp]
\centering
\includegraphics[scale=0.4]{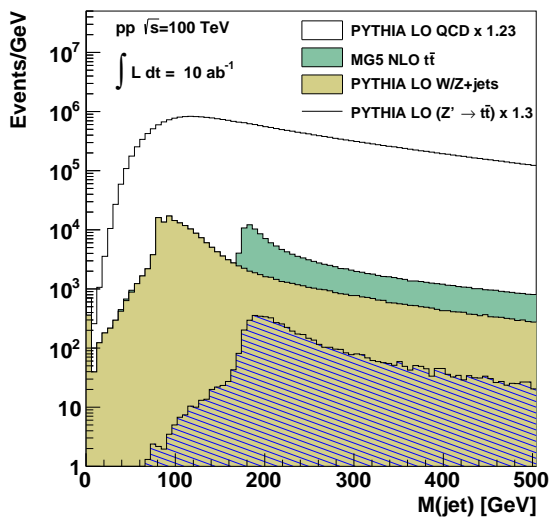}
\caption{Distributions of the leading-jet mass for different processes at a 100~TeV collider.
All background histograms are stacked. The QCD background shown with the white histogram is
dominated by light-flavor dijet events from {\sc Pythia}.
}
\label{h_jet1_mass}
\end{figure}

Figure~\ref{h_jet1_mass} shows the jet masses for background and 
$\zprime\to t\bar{t}$ processes. The signal and SM $t\bar{t}$  processes show a  peak near 170~GeV, 
indicating the complete containment of top-quark decay products within $R=0.5$ jets given the $p_T(t) > 2.5$~TeV requirement. 

\begin{figure}[htp]
\centering
\includegraphics[scale=0.4]{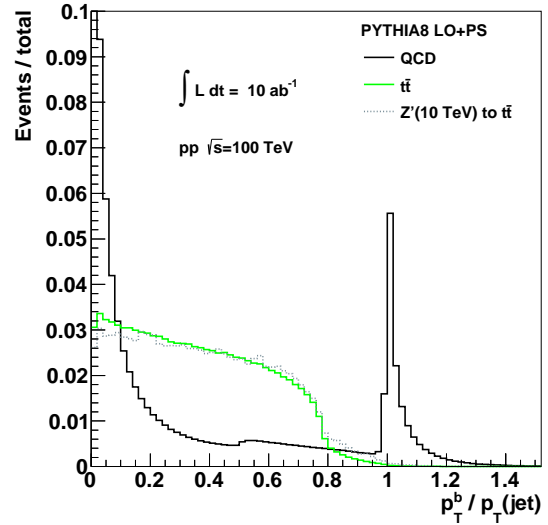}
\caption{The fractional momentum carried by the $b$-quark for non-top and top jets.}
\label{h_bjet1_pt}
\end{figure}

An important part of any analysis that requires reconstruction of top quarks is $b$-tagging. We assume $70\%$ $b$-tagging efficiency, $10\%$ fake rate for $c$-quark jets and $1\%$ fake rate 
 for light-quark (and gluon) jets, similar to that discussed  for Snowmass studies \cite{Auerbach:2013by}.
This choice is motivated by the performance of high-momentum $b$-tagging ($p_T(b)>200$~GeV) at the LHC experiments. 
The $b$-jet is accepted as a boosted-top candidate if a $b$-quark, matched to a jet, has the transverse momentum 
$0.2\cdot p_T(\mathrm{jet})<p_T(b)<0.9\cdot p_T(\mathrm{jet})$.
The motivation for this cut is illustrated in Fig.~\ref{h_bjet1_pt}, which shows
that  $b$-quarks in the parton showers initiated by light-flavor jets are significantly softer
than $b$-quarks from top decays, while the ratio $p_T(b)/p_T(\mathrm{jet})$ exceeds unity for prompt $b$-quarks.  
 
\begin{figure}[htp]
\centering
\includegraphics[scale=0.4]{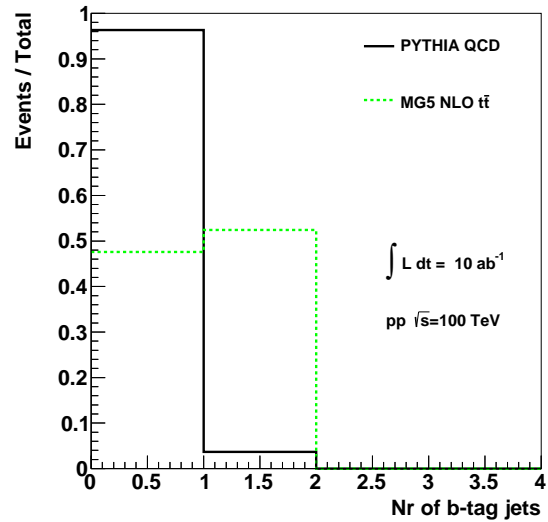}
\caption{The number of b-tags per jet for the SM $t\bar{t}$ events and QCD non-top jet events (from {\sc Pythia8}).}
\label{h_jet1_btag}
\end{figure}

Figure~\ref{h_jet1_btag} shows the fractions of jets that have zero and one $b$-tag. 
The fraction of jets tagged as $b$-jets is less than 70\% for 
$t\bar{t}$ due to the $p_T$ requirements and other inefficiencies of $b$-jet selection.

The $N$-subjettiness characteristics~\cite{Ellis:2009su,Thaler:2010tr}, $\tau_{N}$, of jets has been proposed
as a class of variables with which to study the decay products of a heavy particle inside jets.  $\tau_{N}$ is a measure of the degree to which a jet can be considered as being composed of
 $N$  $k_{T}$-subjets \cite{Thaler:2010tr}. Small values of $\tau_{N}$ correspond to the presence of $N$ or fewer subjets, while large values of $\tau_{N}$ correspond to $>N$ subjets.  
The variable $\tau_{32}$, defined as the ratio of the $N$-subjettiness variables $\tau_3/\tau_2$, is particularly sensitive to hadronically-decaying 
 top-quark initiated jets. 
The variable, $\tau_{21} \equiv \tau_2/\tau_1$ can be used to reject background from $W/Z$ decays.
These variables do not strongly correlate with jet mass and can provide an independent check for the
presence of top quarks. 
The jet substructure variables were obtained by re-running the $k_T$ algorithm over the jet constituents of anti-$k_T$ jets.

\begin{figure}[htp]
\centering
\includegraphics[scale=0.4]{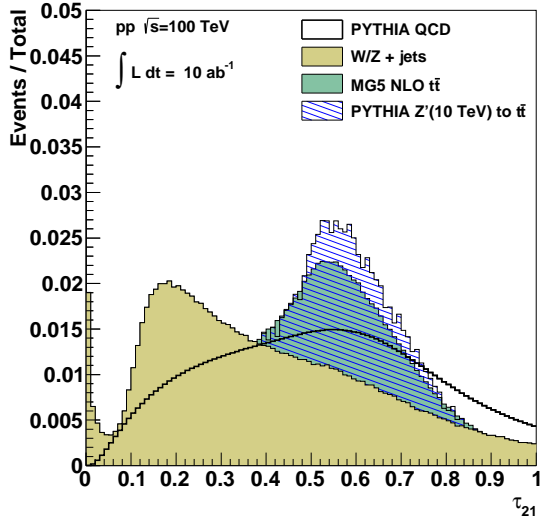}
\caption{Distribution of $\tau_{21}$ for the leading jet, for $\zprime \to t\bar{t}$ and  for major SM processes representing background for the $t\bar{t}$ decay channel.  
}
\label{h_jet_tau21_1}
\end{figure}

\begin{figure}[htp]
\centering
\includegraphics[scale=0.4]{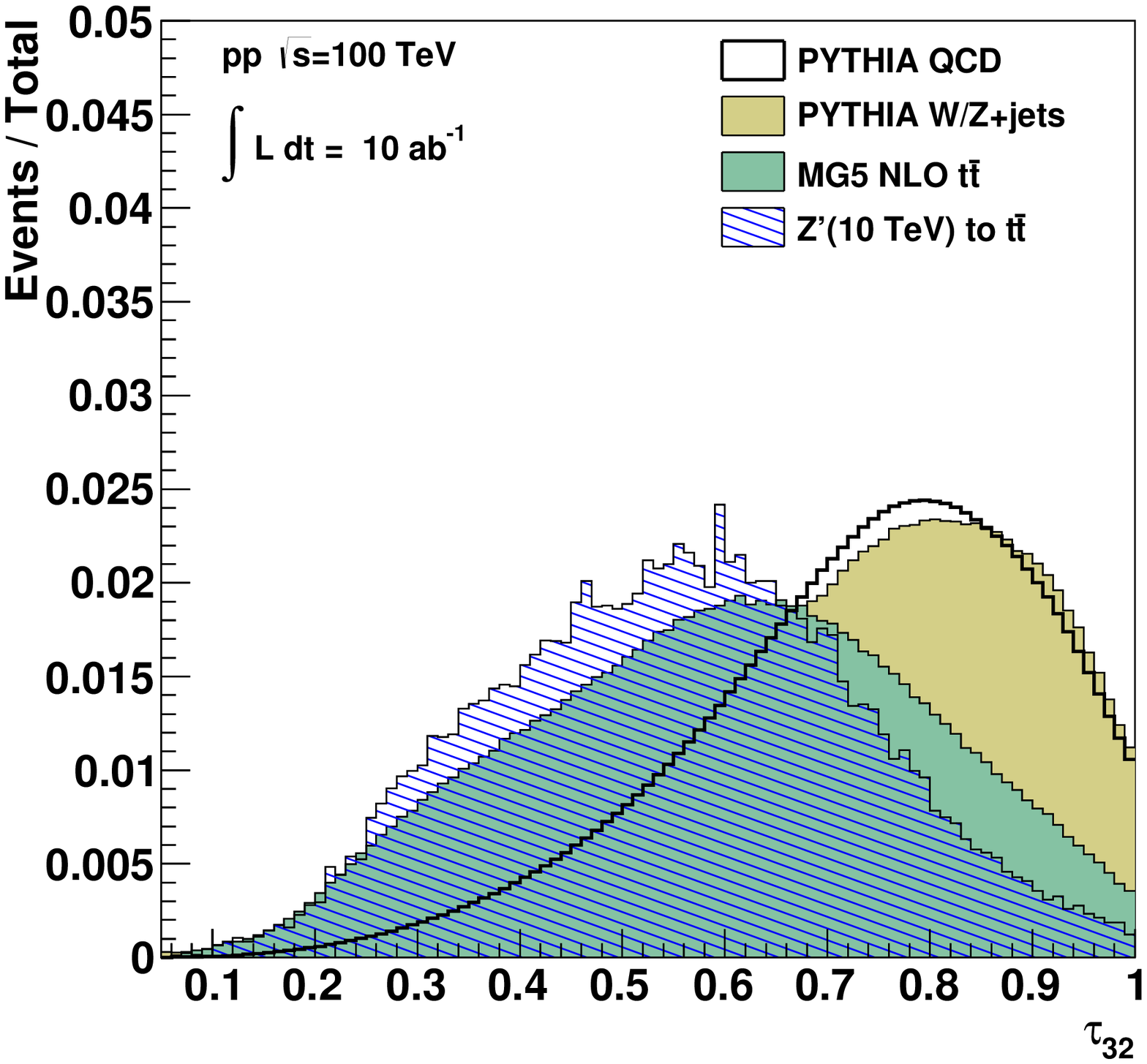}
\caption{
Distribution of $\tau_{32}$ for the leading jet, for $\zprime \to t\bar{t}$ and  for major SM processes representing background for the $t\bar{t}$ decay channel.
}
\label{h_jet_tau32_1}
\end{figure}

Figures~\ref{h_jet_tau21_1} and \ref{h_jet_tau32_1} show the distributions of the $\tau_{21}$ and $\tau_{32}$ variables
for the $\zprime$ signal and SM background processes. These figures show that cutting on these 
variables yields good separation between the signal and the backgrounds. 

\begin{figure}[htp]
\centering
\includegraphics[scale=0.4]{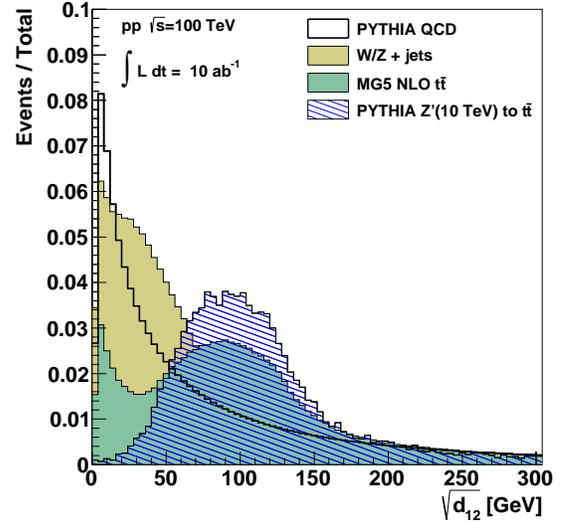}
\caption{
Distribution of $\sqrt{d_{12}}$ for the leading jet, for $\zprime \to t\bar{t}$ and  for major SM processes representing background for the $t\bar{t}$ decay channel.
}
\label{h_jet_d12_1}
\end{figure}

The jet $k_T$ splitting scales \cite{Butterworth:2002tt} can be defined as distance measures 
used to form jets by the $k_T$ recombination 
algorithm \cite{Catani1993187,Ellis:1993tq}. 
This has been extensively studied in Ref.~\cite{ATLAS:2012am}. 
The distribution of the splitting scale $\sqrt{d_{12}}=\min(p_T^1,p_T^2) \times \delta R_{12}$ \cite{ATLAS:2012am} at the final stage of the $k_T$ clustering, where two subjets are merged into the final one,
is shown in Fig.~\ref{h_jet_d12_1}. One can see that the QCD background can be reduced by requiring $\sqrt{d_{12}}>50$~GeV. 

\begin{figure}[htp]
\centering
\includegraphics[scale=0.4]{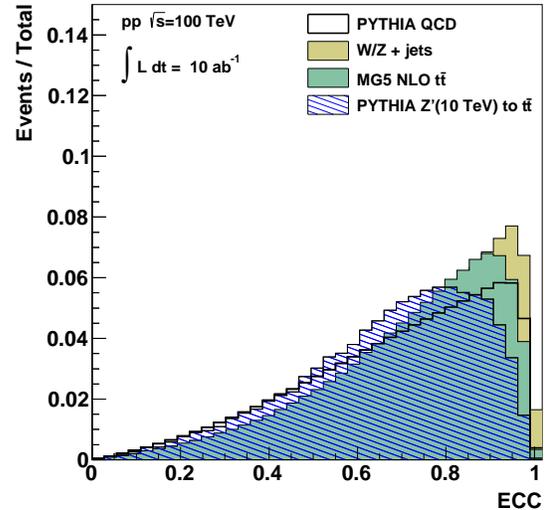}
\caption{
Eccentricity \cite{Chekanov:2010vc} distribution  
of the leading jet for $\zprime \to t\bar{t}$ and  major SM processes representing background for the $t\bar{t}$ decay channel. 
}
\label{h_jet1_ecc}
\end{figure}

The  jet-shape approach based on jet eccentricity \cite{Chekanov:2010vc} is another method that has potential to reduce QCD background processes.  The jet eccentricity 
is less sensitive to jet substructure as
no attempt is made to resolve kinematic characteristics
of separate subjets inside jets.
Figure~\ref{h_jet1_ecc} shows the jet eccentricity (ECC) for leading jets, without
applying other cuts to enhance the $\zprime$ signal. 
It can be seen that the eccentricity cut at 0.9 rejects some QCD-dijet and $W/Z$ background.

\begin{figure}[htp]
\centering
\includegraphics[scale=0.4]{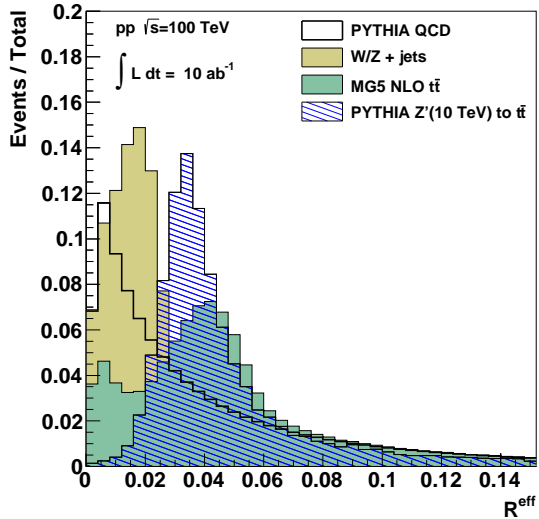}
\caption{
Jet effective radius  
of the leading jet for $\zprime \to t\bar{t}$ and  major SM processes representing background for the $t\bar{t}$ decay channel.
}
\label{h_effR_1}
\end{figure}

We also consider the effective radius ($R^{\rm eff}$) of the leading jet. 
The effective radius is the average of the energy weighted radial distance in $\eta-\phi$ space of jet constituents. Jets with only soft splitting are most energetic along the jet axis compared to jets with decays from heavy particles which have an intrinsic $k_{T}$ related to the decay particle's mass.
The distribution of the effective radius for  $\zprime \to t\bar{t}$ and background processes
is shown in Fig.~\ref{h_effR_1}. This figure shows a good separation power of $R^{\rm eff}$ to reject QCD background.

As a  check of these discriminating variables using an alternate QCD background generator, {\sc Pythia8} 
 was replaced with the {\sc Herwig++} generator for dijet QCD processes. 
The latter contains an alternative parton shower and hadronisation model. 
In particular, {\sc Herwig++} includes the production of $W/Z$ bosons inside the parton shower,
  which has the potential to be an irreducible background for the substructure and jet-shape variables.
We have found  that all the conclusions on background reduction obtained with  {\sc Pythia8}
 still hold with {\sc Herwig++}.

\begin{figure}[htp]
\centering
\includegraphics[scale=0.4]{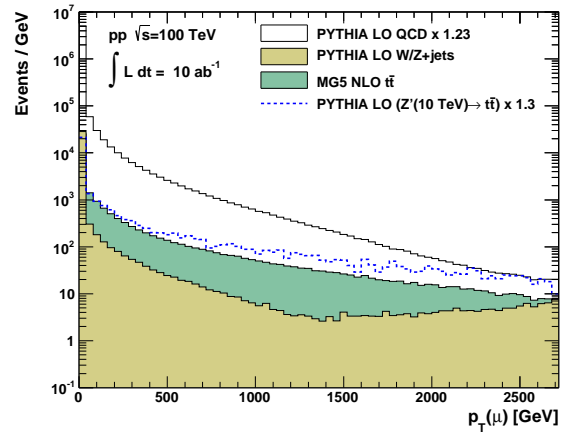}
\caption{
Transverse-momentum distribution of muons leading in $p_T$, in the hemisphere 
opposite the leading jet.  The background histograms are stacked. 
}
\label{h_ptmuon2}
\end{figure}

Finally, $t\bar{t}$ production is characterized by the presence of high-$p_T$ leptons in the case of leptonic $W$ decays. 
As example, Figure~\ref{h_ptmuon2} shows  the transverse momentum of the highest-$p_T$ muon opposite to 
the leading jet,  where the latter is typically 
due to fully-hadronic top decays. A selection cut at $1-1.5$~TeV can be applied to reject the SM background.
However, such a cut substantially reduces the statistics.  

The event shapes discussed above have certain degrees of correlations.
The  correlations can be studied in terms of the correlation coefficient, $\rho$, that represents a degree of linear dependence between two variables. 
The correlation coefficient varies from -1 (perfect negative correlation) to +1 (perfect positive correlation).
The largest positive correlation of $\rho=0.66$ was observed between jet masses and the splitting scale $\sqrt{d_{12}}$. 
Jet mass also correlates with $R^{\rm eff}$ and the eccentricity ($\rho=0.4$). 
The correlation between $\tau_{32}$ and $\sqrt{d_{12}}$ is $\rho=-0.35$.
Other variables typically have correlations with $|\rho|<0.3$.

\begin{figure}[htp]
\centering
\includegraphics[scale=0.4]{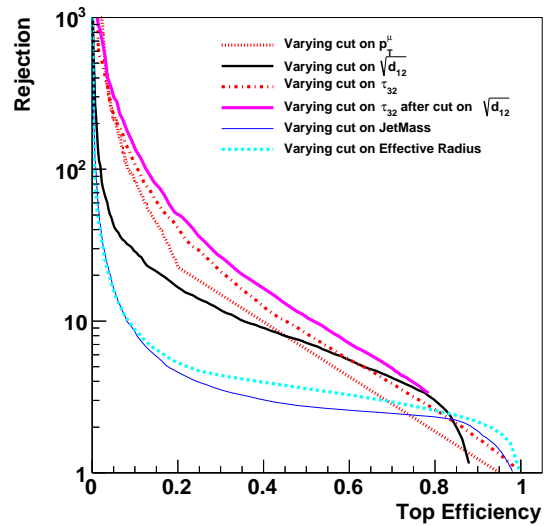}
\caption{
Rejection factor for QCD jets versus the 
efficiency of reconstruction of top quarks for different variables used to select top jets.
}
\label{h_erdmann}
\end{figure}

Figure~\ref{h_erdmann} illustrates the rejection factor for QCD background events as a function of the efficiency
of top-quark reconstruction. It can be seen that cuts on the 
jet masses and $R^{\rm eff}$  are not very effective in rejection of QCD background,
given their low efficiency in selecting top quarks. A cut on the muon $p_T$  is an attractive option, 
but it can only be used 
for sufficiently large luminosity.  
The $b$-tagging performance is not shown on this plot since we assume a fixed operating point for the $b$-tagging efficiency ($70\%$).
We will return to the discussion of how to optimize selection cuts in the following sections.

\section{Top jets}

In this section we will illustrate the effectiveness of using jet substructure variables and $b$-tagging
in selecting jets initiated by top quarks. 
The use of the jet discriminating variables can help to reconstruct 
individual ``top-tagged jets'', i.e. jets that are consistent with being  initiated by top quarks but do not necessarily arise from the $t\bar{t}$ process, as the usual 
requirement to identify a second top (anti-top) is not imposed \cite{Auerbach:2013by,Calkins:2013ega}.
This can be crucial to identify high $p_T$ SM single-top processes, or new processes that do not possess the signatures of 
the $t\bar{t}$ event topology such as the decay of new $W'$ bosons. 

\begin{figure}[htp]
\centering
\includegraphics[scale=0.4]{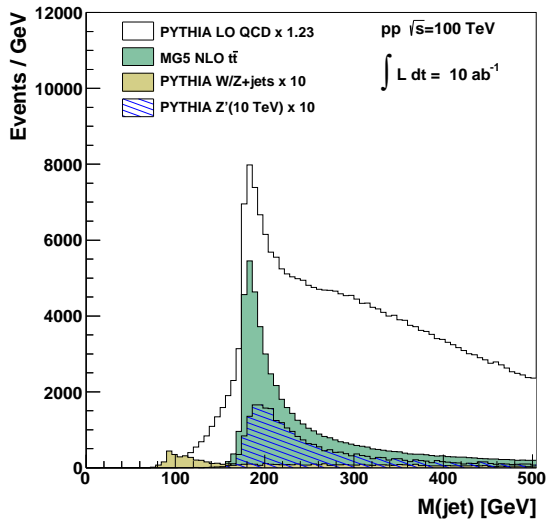}
\caption{Jet masses after $b$-tagging and the cuts on $\tau_{32}$ and $\tau_{21}$.
No requirements on the presence of a second top (anti)quark were imposed. The contributions of $W/Z$ and $\zprime$ processes are multiplied by the 
factor ten to increase visibility. All background histograms are stacked.  
}
\label{h_mass_all}
\end{figure}

Figure~\ref{h_mass_all} shows the jet masses after $b$-tagging and the requirements 
$\tau_{32}<0.7$, $0.3<\tau_{21}<0.8$ and $\sqrt{d_{12}}>50$~GeV 
on a single jet, without requiring the presence of a second $t(\bar{t})$ quark decay.
The result shows that the $t\bar{t}$ process  dominates the 170~GeV jet mass region.
The figure also shows the contribution of top jets from the $\zprime$ process (multiplied by a factor 10 for better visibility).
The top-quark requirements also select boosted $W/Z$ jets (their contribution in Fig.~\ref{h_mass_all} is also scaled by 10).
The eccentricity and the effective jet radius were also studied,
but no significant change for Fig.~\ref{h_mass_all} was found due to their correlations with other variables.

We conclude that selecting single jets using $b$-tagging and jet substructure variables 
is an attractive possibility for a 100~TeV collider, since such single-jet selection is sensitive to events beyond
the standard $t\bar{t}$  event topology.

\section{Dijets after top-quark tagging}

\begin{figure}
\centering
  \subfigure[Dijet mass distribution with  $\zprime \to t\bar{t}$ signal]{
  \includegraphics[scale=0.45, angle=0]{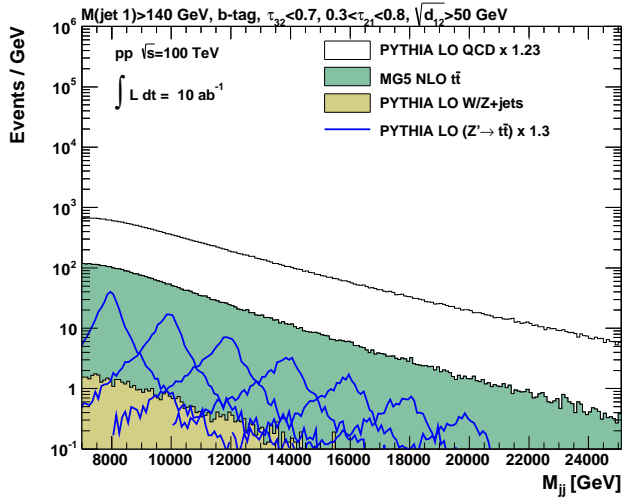}
  }
  \subfigure[Dijet mass distribution with $\gkk \to t\bar{t}$ signal]{
  \includegraphics[scale=0.45, angle=0]{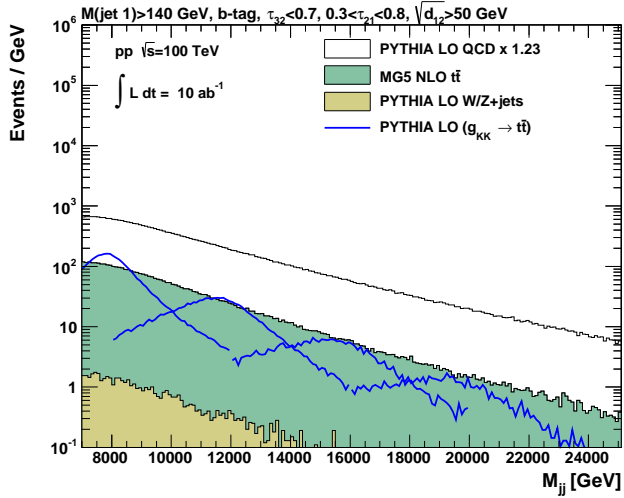}
  }
\caption{The distributions of dijet mass after the requirements  
$M(\mathrm{jet})>140$ GeV, $\tau_{32}<0.7$,  $0.3<\tau_{21}<0.8$, $\sqrt{d_{12}}>50$~GeV 
and single $b$-tagging.
No requirements on the second jet were imposed.
The expectation for $\zprime$ and $\gkk$ processes are shown with the lines. The background histograms are stacked. 
The QCD background shown with the white histogram is 
dominated by light-flavor dijet events from {\sc Pythia}.
}
\label{h_jetjet3}
\end{figure}

Figure~\ref{h_jetjet3} shows the jet masses after single $b$-tagging and 
the cuts $M(\mathrm{jet})>140$~GeV,  $\tau_{32}<0.7$, $0.3<\tau_{21}<0.8$ and $\sqrt{d_{12}}>50$~GeV applied to the leading jet.
The figure shows that the  signal-over-background ratio  for $\zprime$ and $\gkk$ processes is significantly improved (see
Fig.~\ref{h_jetjet_zprime0} for comparison). 

\begin{figure}
\centering
  \subfigure[Dijet mass distribution with  $\zprime \to t\bar{t}$ signal]{
  \includegraphics[scale=0.45, angle=0]{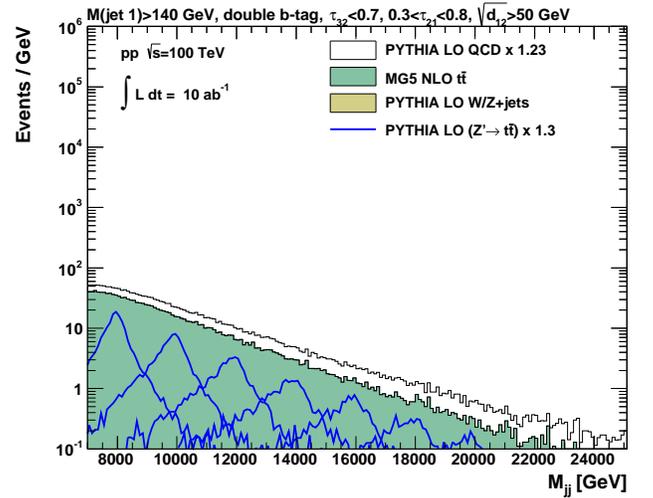}
  }
  \subfigure[Dijet mass distribution with $\gkk \to t\bar{t}$ signal]{
  \includegraphics[scale=0.45, angle=0]{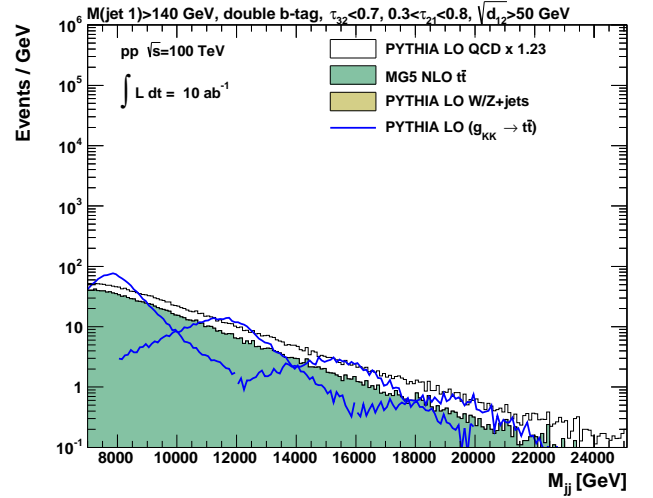}
  }
\caption{Same as Fig.~\ref{h_jetjet3}, but after  the 
$b$-tagging requirement on both jets.
No subjet requirements were imposed 
 on the second jet.  
The expectations  for $\zprime$ and $\gkk$ processes are shown with the lines. The background histograms are stacked. 
}
\label{h_jetjet5}
\end{figure}

Similarly, Figure~\ref{h_jetjet5}  shows
the dijet-mass distributions after double $b$-tagging, i.e. when both jets are required to pass
the $b$-tagging requirement. It can be seen that the S/B ratio increases by a factor 130, compared to
the case without cuts shown in Figs.~\ref{h_jetjet_zprime0} and \ref{h_jetjet_kkgluon0}. 
However, after applying these selection requirements, 
there is a substantial decrease in statistics.  

Selection cuts based on jet eccentricity and high-$p_T$ muons were studied, but such cuts did not show large improvement in the signal-over-background ratio.   
After applying the cuts on $b$-tagging and the jet substructure, an additional cut on muon $p_T$ did not yield significant further reduction in the SM background.

To illustrate the optimization of the selection cuts, 
Tables~\ref{tab_cuts1} and \ref{tab_cuts2} show the signal-over-background ratios in the dijet mass window 
9--11~TeV for a $\zprime$ boson with a nominal mass of 10~TeV.
The tables show that $b$-tagging has the most significant impact on the S/B.
Table~\ref{tab_cuts2} illustrates that one promising option is to 
use a selection based on double $b$-tagging and jet-substructure variables applied  for one jet, which leads
to S/B$\simeq 0.2$, while still retaining good statistics assuming an integrated luminosity of 10 ab$^{-1}$.
 
\begin{table}[h]
\begin{tabular}{|c|c|c|c|c|}
No cuts & JS     & $b$-tag  & $b$-tag+JS & $b$-tag+JS+$\mu$ \\ \hline 
0.0007  & 0.0025 & 0.013     & 0.025      &  0.12           \\   
\end{tabular}
\caption{The signal-over-background ratios for $\zprime$ with the mass 10~TeV for different combination
of the selection cuts. The  S/B ratio was calculated in the mass window 9--11~TeV.
The background includes all the SM processes, such as QCD light-flavor and $b \bar{b}$, $t\bar{t}$ and $W/Z$ production. 
The abbreviation "JS" means a selection based 
on jet-shape and jet-substructure variables only, i.e. $M(\mathrm{jet})>140$~GeV, $\tau_{32}<0.7$, 
$0.3<\tau_{21}<0.8$, $\sqrt{d_{12}}>50$~GeV. 
The symbol $\mu$ indicates a requirement of $p_T(\mu)>1.2$~TeV. 
}
\label{tab_cuts1}
\end{table}

\begin{table}[h]
\begin{tabular}{|c|c|c|c|c|c|}
No cuts & JS2    &   $b$-tag &  $b$-tag+JS1 & $b$-tag+JS2 & $b$-tag+JS1+$\mu$ \\ \hline
0.0007  & 0.007 &   0.16     &  0.19                 & 0.21                 & 0.36 \\
\end{tabular}
\caption{Same as Table~\protect{\ref{tab_cuts1}}, but the $b$-tagging and jet substructure requirements are applied to both jets.
The abbreviation "JS1" indicates the jet substructure cuts for a single jet, while "JS2" indicates the  application of
the jet-substructure  requirements for both jets.
Although the S/B ratio is the largest for the last column, the statistics are not sufficient to obtain 
a competitive 95\% CL sensitivity compared to other selections.
}
\label{tab_cuts2}
\end{table}

\section{Signal Sensitivity}

\begin{figure}[htp]
\centering
\includegraphics[scale=0.45]{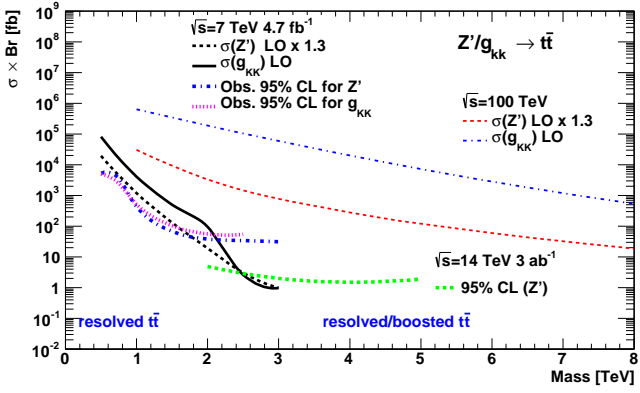}
\caption{{\sc Pythia8} $\sigma \times$Br for $\zprime /\gkk$ processes for  $pp$ collisions at 100~TeV.  Also shown is a compilation of $\zprime /\gkk$ cross sections and exclusion limits 
for $pp$ collisions at lower energies,
7~TeV \cite{Aad:2013nca} and 14~TeV \cite{Agashe:2013fda}.
}
\label{exclu_summary_pub}
\end{figure}

Figure~\ref{exclu_summary_pub} shows 
the $\sigma \times$Br for $\zprime /\gkk$ processes  calculated using the {\sc Pythia8} generator 
for $pp$ collisions at 100~TeV.
As before, the $\zprime$ cross section includes the $k-$factor.
The figure also shows 
a compilation of exclusion limits for heavy $\zprime /\gkk$ particles decaying to
$t\bar{t}$. 
This compilation is based on the studies of  a $t\bar{t}$ resonance in the 
lepton+jets final state \cite{Aad:2013nca}.
Other studies \cite{Chatrchyan:2012ku}
show similar limits. 
Figure~\ref{exclu_summary_pub} shows the Snowmass studies \cite{Agashe:2013fda} based 
on a fast detector simulation (assuming no pileup) \cite{deFavereau:2013fsa} for $14$~TeV $pp$ collisions.
It should be pointed out that the studies of collisions at 7 and 14~TeV were done using a combination of boosted 
and resolved techniques, since the boost was not as large compared to the situation discussed in this paper. 

\begin{figure}[htp]
\centering
\includegraphics[scale=0.45]{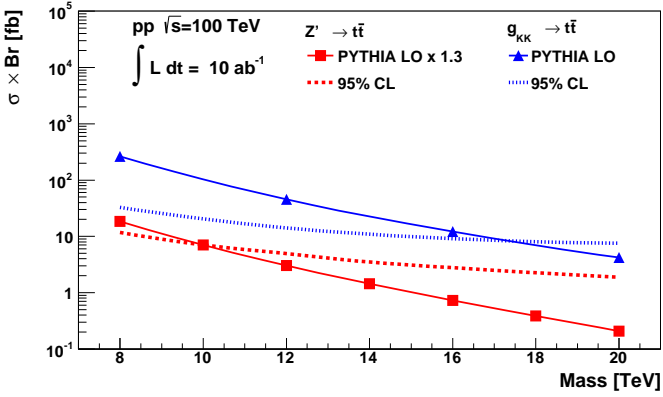}
\caption{A compilation  of sensitivities for $\zprime$ and  $\gkk$ bosons decaying to $t\bar{t}$  at a 100 TeV $pp$ collider using the ``fully-boosted'' regime without resolving separate decay products of top quarks. 
The calculations for $\sigma \times$Br were performed using LO QCD, with a $k$-factor of 1.3 for $\zprime$ production assumed
from lower-energy colliders. The sensitivities are given  after the selection
$M(\mathrm{jet})>140$~GeV, $\tau_{32}<0.7$,  $0.3<\tau_{21}<0.8$, $\sqrt{d_{12}}>50$~GeV   
and single $b$-tagging, assuming $10$~$\abb$ of integrated luminosity.
}
\label{exclu_summary_7}
\end{figure}

\begin{figure}[htp]
\centering
\includegraphics[scale=0.45]{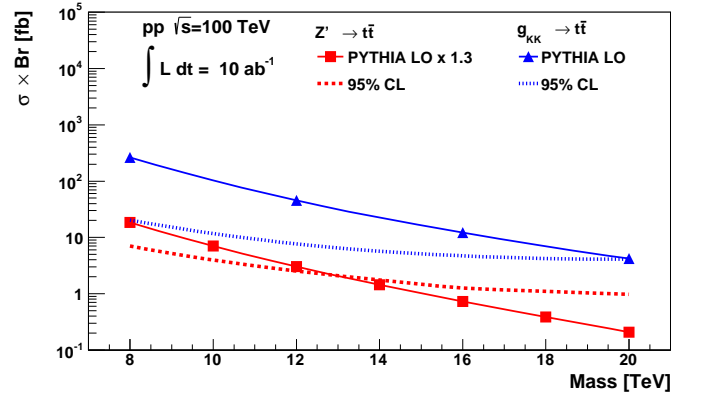}
\caption{Same as Fig.~\ref{exclu_summary_7}, but using double $b$-tagging.
}
\label{exclu_summary_5}
\end{figure}

The current study 
extends the sensitivities shown in Fig.~\ref{exclu_summary_pub} to masses above $8$~TeV using ``fully-boosted'' reconstruction of
$t\bar{t}$ at a 100~TeV $pp$ collider.  
Figure~\ref{exclu_summary_7} shows the production cross section
and the obtained sensitivities for $\zprime/\gkk$ resonances for a 100~TeV collider assuming
$10$~$\abb$ of integrated luminosity. 
We use variables designed to increase the signal-over-background
ratio  as discussed in the previous section and shown in Fig.~\ref{h_jetjet3}, i.e. 
$M(\mathrm{jet})>140$~GeV, $\tau_{32}<0.7$,  $0.3<\tau_{21}<0.8$, $\sqrt{d_{12}}>50$~GeV  
and single $b$-tagging.
Figure~\ref{exclu_summary_7} can 
directly be compared to the sensitivities shown in Fig.~\ref{exclu_plot0}, which were obtained without any selection cuts.
Figure~\ref{exclu_summary_5} shows the cross sections and the sensitivities after using double $b$-tagging as applied for 
Fig.~\ref{h_jetjet5}. 
Table~\ref{tbl:mubd} shows the values of $\sigma \times$Br for theory and experimental sensitivity as a function of resonance mass
used in \ref{exclu_summary_5}.
It can be seen that a 100~TeV collider has the sensitivity for heavy particles decaying to $t\bar{t}$,
with mass up to 20~TeV. The best sensitivity is obtained using double $b$-tagging. 
The assumed $10$~$\abb$ of integrated luminosity
is also sufficient to be sensitive to a $\zprime$ boson  with mass up to 13~TeV. 
 Using single $b$-tagging, the sensitivity for $\gkk$ ($\zprime$) extends to  masses of 17 (10)~TeV.
Note that this statement is  valid for the default {\sc Pythia8} model 
that calculates the production cross sections using LO QCD (after assuming the  $k$-factor of 1.3  for $\zprime$ production).

\begin{table}[!ht]
\begin{ruledtabular}
\begin{tabular}{lcccc}
 mass & \multicolumn{4}{c}{$\sigma \times$Br (fb)} \\
(TeV)    & $\zprime$ (th.)  & $\zprime$ (exp.) & $\gkk$ (th.) & $\gkk$ (exp.)  \\
\hline
8  & 18.46 & 7.00  & 262.3 & 20.2 \\
10 & 7.03 & 3.97  & & \\
12 & 3.02 & 2.54 & 45.4 & 7.7 \\
14 & 1.44 & 1.75 & & \\
16 & 0.73 & 1.27 & 12.2 & 4.7 \\
18 & 0.39 & 1.10 & &  \\
20 & 0.21 & 0.98  & 4.2 & 4.1 \\
\end{tabular}
\end{ruledtabular}
\caption{Values of $\sigma \times$Br for theory and experimental sensitivity as a function of resonance mass shown in Fig.~\ref{exclu_summary_5}.}
\label{tbl:mubd}
\end{table}

The 95\% CL sensitivity estimates  for a 100~TeV collider with the integrated luminosity of $10$~$\abb$ 
are rather general, as long as the widths of the $t\bar{t}$ resonances are similar to  those discussed in this paper.
The $\sigma \times$Br sensitivity is close to 2-4~fb in the mass range 15-20~TeV
for heavy particles with widths similar to $\zprime$ or $\gkk$.
It was shown that the usage of  $b$-tagging and the boosted-top techniques can increase the 
signal-over-background ratio 
for heavy states decaying to $t\bar{t}$
by more than a factor of two hundred, and increase
the sensitivity by a factor 10.
While the ``resolved'' method \cite{Aad:2013nca,Chatrchyan:2012ku} for top reconstruction yields greater improvements with these techniques, 
it should be emphasized that we are
dealing with an especially difficult boosted topology when $t\bar{t}$ events lead to back-to-back  high-$p_T$ jets.

It is useful to estimate how the sensitivity would improve with integrated luminosity. 
Since there is significant SM background under the $\zprime$ and $\gkk$ mass peaks,
the sensitivity is expected to scale as $S/\sqrt{B}$ which increases 
as the square root of the integrated luminosity.  
For integrated luminosities of 30~ab$^{-1}$ and 150~ab$^{-1}$ respectively,
the $\zprime$ mass reach would increase from 13~TeV to 16~TeV and 19~TeV respectively, 
using the selection criteria mentioned in Fig.~\ref{exclu_summary_5}. 
If the selection criteria mentioned in Fig.~\ref{exclu_summary_7} 
were used, these higher integrated luminosities would increase the $\zprime$ mass reach from 10~TeV to 12~TeV and 16~TeV respectively.
With 30~ab$^{-1}$ of integrated luminosity, the $\gkk$ mass reach would increase from 17~TeV to 19.5~TeV.

\section{Conclusion}
The sensitivity to $\zprime$  and $\gkk$ bosons in the mass range of 8-20~TeV 
decaying to $t\bar{t}$ is discussed for a 100 TeV $pp$ collider.
It was illustrated how several popular discriminating variables 
can be used to reduce the background for searches
for heavy particles decaying to highly boosted top quarks.
The discriminating variables can increase the sensitivity on the $\sigma \times$~Br of 
$\zprime$  and $\gkk$ bosons by a factor 10, leading to 
 95\% CL signal sensitivity values of $\sigma \times$Br of 2-4~fb depending on the resonance width.
 The combined use of $b$-tagging and jet substructure variables 
can increase 
the signal-over-background ratio 
for heavy states decaying to $t\bar{t}$ 
by more than a factor two hundred. 

This study shows that a 100~TeV collider  with an integrated luminosity of 10~$\abb$  can be sensitive 
to a $\gkk$ resonance with the mass 20~TeV, assuming 
the LO QCD cross section for the $\gkk$ production. 
The study indicates that the assumed integrated luminosity
is also sufficient to be sensitive to $\zprime\to t\bar{t}$ decays with mass of $13$~TeV, 
a channel that is very challenging compared to the di-lepton decays of $\zprime$ bosons. 
It is important to check this conclusion after incorporating a realistic detector simulation.
The paper suggests several important criteria for a future 100~TeV experiment:
a highly efficient $b$-tagging and the capability of resolving substructure in jets of $R=0.5$ with 
transverse momenta above $3$~TeV.

It should be pointed out that  the sensitivities presented in this paper should be considered in a broader context
 since they can illustrate  the physics reach of a 100~TeV collider to  generic heavy particles,
assuming the most essential selection cuts needed to reconstruct  highly-boosted top quarks from $t\bar{t}$ resonances. 
These results  can be useful to illustrate the capability of a 100~TeV collider and to develop
theoretical models.

\section*{Acknowledgments}
We thank Lily Asquith for a discussion of the implementation of the splitting scale variable.
We also would like to thank Michelangelo Mangano for checking NLO calculations for $t\bar{t}$. 
The submitted manuscript has been created by UChicago Argonne, LLC,
Operator of Argonne National Laboratory (``Argonne'').
Argonne, a U.S. Department of Energy Office of Science laboratory,
is operated under Contract No. DE-AC02-06CH11357.
A fraction of the simulated event samples presented in this paper  were 
generated using the ATLAS Connect virtual cluster service.
This research used resources of the Argonne Leadership Computing Facility at Argonne National Laboratory, which is supported by the Office of Science of the U.S. Department of Energy under contract DE-AC02-06CH11357.

\clearpage
\bibliography{biblio}

\newpage

\appendix

\begin{widetext}
\onecolumngrid

\section*{Appendix}

\begin{figure}
\centering
\includegraphics[scale=0.4, angle =90]{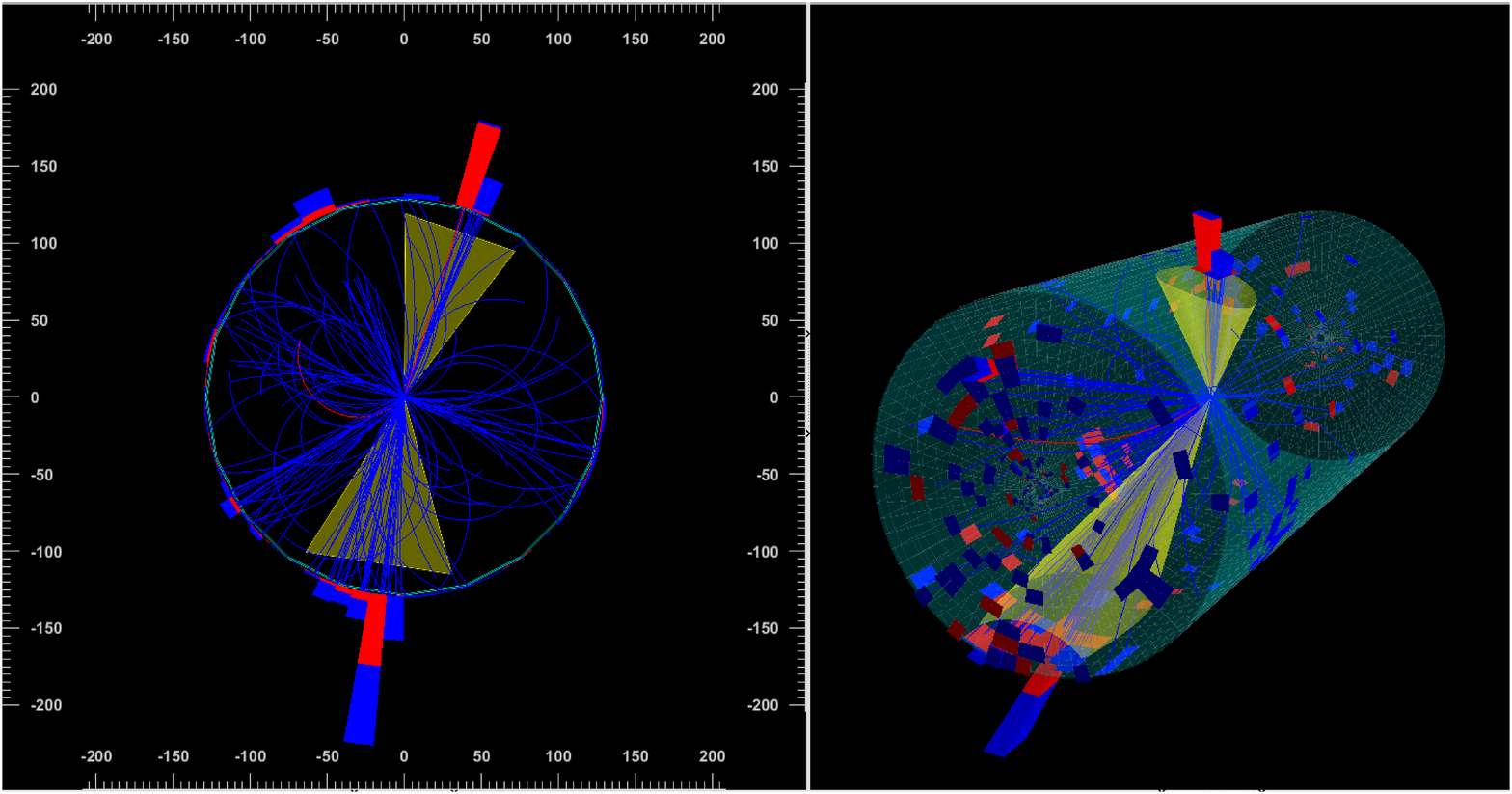}
\caption{An event display of a typical $\zprime$(M=10~TeV) decaying to $t\bar{t}$ at
a $100$~TeV collider.
Two jets with transverse momenta above 3~TeV are shown with the yellow cones. The jets are
reconstructed using the anti-$k_T$ algorithm \cite{Cacciari:2008gp}
with  a distance parameter of 0.5 using the {\sc FastJet} package~\cite{fastjet}.
The event display was created using the {\sc Delphes}
fast simulation \cite{deFavereau:2013fsa}, 
{\sc HepSim} \cite{Chekanov:2014fga} and the Snowmass detector setup \cite{Anderson:2013kxz}.
The blue lines show charged hadrons and red lines  show contributions from electrons.
See the text for details.
}
\label{View2}
\end{figure}
\end{widetext}

\end{document}